\documentclass[prb,twocolumn,showpacs,amsmath,amssymb]{revtex4}

\usepackage{graphicx}
\usepackage{dcolumn}
\usepackage{bm}

\begin{document}


\title{Magnetic properties of the Haldane-gap material NENB}

\author{E. \v{C}i\v{z}m\'{a}r}
\email{e.cizmar@fzd.de}
\author{M. Ozerov}
\author{O. Ignatchik}
\author{T. P. Papageorgiou}
\author{J. Wosnitza}
\author{S. A. Zvyagin}
\affiliation{Dresden High Magnetic Field Laboratory (HLD),
Forschungszentrum Dresden-Rossendorf, 01314 Dresden, Germany}
\author{J. Krzystek}
\affiliation{National High Magnetic Field Laboratory, Florida
State University, Tallahassee, FL 32310, USA}
\author{Z. Zhou}
\affiliation{Department of Physics and Astronomy, Wayne State University,
Detroit, MI 48201, USA}
\author{C. P. Landee}
\author{B. R. Landry}
\author{M. M. Turnbull}
\affiliation{Department of Physics and Carlson School of Chemistry, Clark
University, Worcester, MA 01060, USA}
\author{J. L. Wikaira}
\affiliation{Department of Chemistry,
University of Canterbury, Christchurch, New Zealand}

\date{\today}

\begin{abstract}
Results of magnetization and high-field ESR studies of the new
spin-1 Haldane-chain material
[Ni(C$_2$H$_8$N$_2$)$_2$NO$_2$](BF$_4$) (NENB) are reported. A
definite signature of the Haldane state in NENB was obtained. From
the analysis of the frequency-field dependence of magnetic
excitations in NENB, the spin-Hamiltonian parameters were calculated,
yielding $\Delta = 17.4~K$, $g_{\parallel}=2.14$, $D=7.5~K$, and $|E|=0.7~K$ for the Haldane gap, $g$
factor and the crystal-field anisotropy constants, respectively. The
presence of fractional $S=1/2$ chain-end states, revealed by
ESR and magnetization measurements, is found to be
responsible for spin-glass freezing effects. In addition, extra
states in the excitation spectrum of NENB have been observed in
the vicinity of the Haldane gap, which origin is discussed.

\end{abstract}

\pacs{76.30.-v, 74.25.Ha, 75.50.Gg, 75.30.Gw}
\maketitle

Antiferromagnetic (AFM) quantum spin-1 chains have been the
subject of intense theoretical and experimental studies, fostered
especially by the Haldane conjecture.\cite{haldane} In accordance to the valence-bond-solid (VBS) model proposed by Affleck \textit{et
al.}\cite{VBS}, each $S=1$ spin can be regarded as a symmetric
combination of two $S=1/2$ moments forming a spin-singlet ground
state. The value of the energy gap $\Delta=0.41J$ (where $J$ is the
spin-spin exchange coupling) has been estimated for an  $S=1$
isotropic Heisenberg AFM chain using the density-matrix
renormalization-group (DMRG)\cite{DMRGgap} and exact
diagonalization\cite{exactgap} techniques. The presence of the
Haldane gap was experimentally revealed in a number of materials (see for instance Ref. 
\onlinecite{RenardNENP,LuNENP,YBaNiO1,YBaNiO2}). Nowadays, investigations of intriguing
properties of Haldane materials (particular their
high-field magnetic properties) continue to attract a great deal of
attention. For instance, experimental studies of the Haldane materials
NDMAP\cite{NDMAP1,NDMAP2} and TMNIN\cite{TMNIN} suggest the
realization of the Tomonaga-Luttinger-liquid high-field
phase.\cite{TLL1, TLL2} At sufficiently low temperature, an applied magnetic
field can induce long-range magnetic order, which can be
effectively described as the condensation of a dilute gas of
magnons.\cite{bec1,bec2}

In  this paper, we utilize 
magnetization and electron spin resonance (ESR) techniques to study the new spin-1 Haldane  compound
[Ni(C$_2$H$_8$N$_2$)$_2$NO$_2$](BF$_4$) (abbreviated as NENB). The
presence of the Haldane state in NENB has been confirmed
experimentally. Analysis of the frequency-field
dependence of magnetic excitations in NENB allowed us to determine
the spin-Hamiltonian parameters, yielding $g_{\parallel}=2.14$,
$D=7.5~K$, and $|E|=0.7~K$ for the $g$ factor and the crystal-field
anisotropy constants, respectively. The presence of spin-1/2
excitations in nominally pure NENB crystals revealed by our experimental observations suggested the existence
of weakly-interacting antiferromagnet droplets originating from
broken spin bonds. In addition, extra states in the
excitation spectrum in the vicinity of the Haldane gap were observed, which
origin is discussed.

\begin{figure}[b]
\includegraphics[width=0.45\textwidth,clip=true]{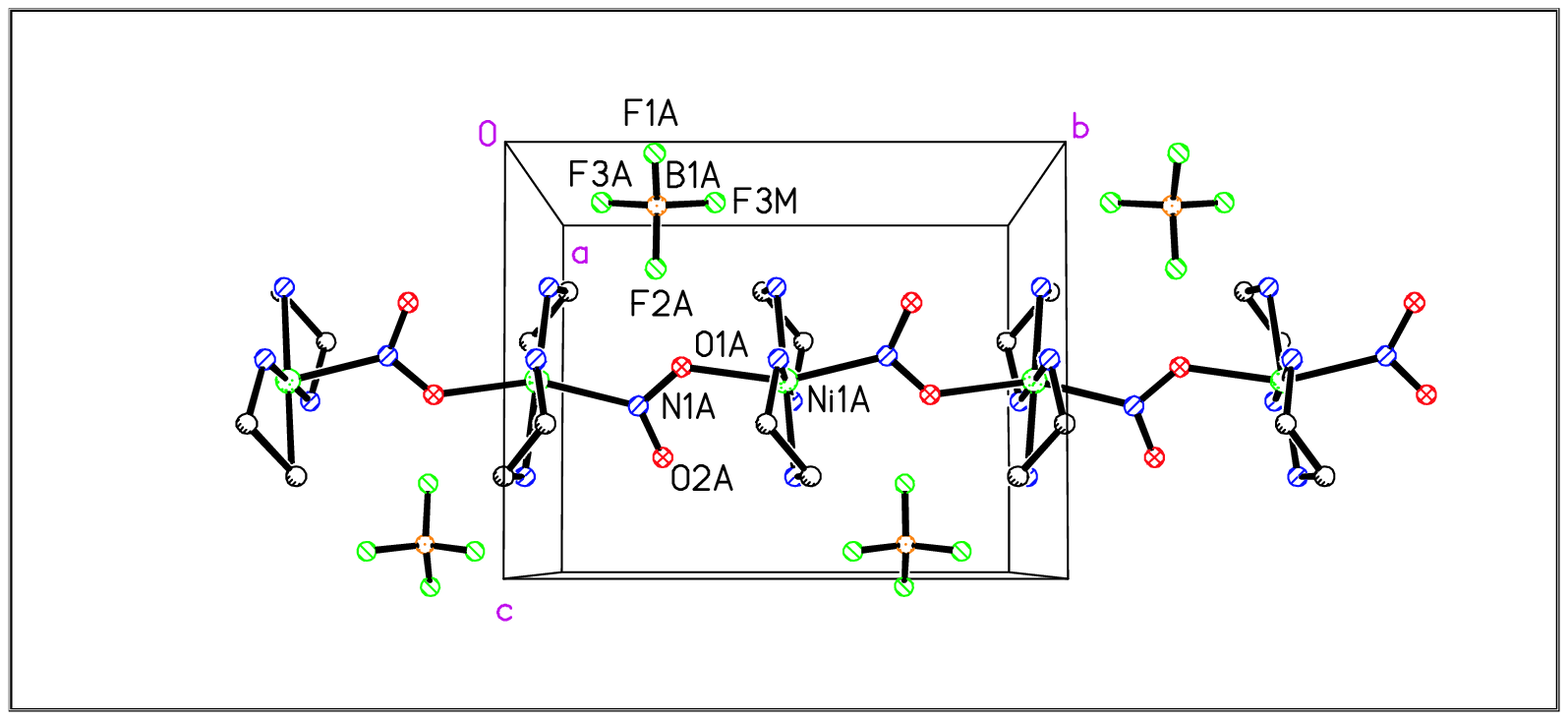}
\caption{\label{fig:structure} (Color online) The structure of
NENB showing the polymeric chain arrangement, the orientation of
the nitrite ligands, and the placement of the $BF_4$ anions
relative to the chain. Atoms are shown as spheres of arbitrary
size and hydrogen atoms are not shown for clarity.}
\end{figure}

Single crystals of NENB were grown by the reaction of
[Ni(C$_2$H$_8$N$_2$)$_3$](BF$_4$)$_2$ with Ni(BF$_4$)$_2
{\cdot}$6H$_2$O and NaNO$_2$ in aqueous solution. Careful, slow
evaporation in a partially covered container yielded ruby-red
crystals of up to 3-4~mm in length and 1-2~mm$^2$ cross section.
Single-crystal X-ray analysis showed an orthorhombic unit cell,
space group $Pnma$ (similar to that of the isostructural compound NENP
\cite{structNENP}), with $a=15.0373(5)$~\AA, $b=10.2276(3)$~\AA, and
$c=7.9719(2)$~\AA. Each Ni$^{2+}$ ion is pseudo-octahedrally
coordinated. The four nitrogen atoms of the two ethylenediamine
rings (ethylenediamine~=~C$_2$N$_2$H$_8$) make a slightly distorted
square planar symmetry with bridging
NO$^{-}_{2}$ ions creating the octahedral axis approximately
parallel to the chain direction. The individual chains are well isolated by the inorganic counterions BF$^{-}_{4}$. As follows from
crystallographical analysis the mean planes of successive
Ni$^{2+}$ units are canted by an angle of 14.1$^{\circ}$ to each other. The orientation of the nitrite ions within a single
chain appears to be uniform (one N atom and one O atom are coordinated
to each Ni$^{2+}$ center) as shown in Fig.\ \ref{fig:structure} 
(full crystallographic details are available from the
Cambridge Crystallographic Data Center, deposition number 637023).\cite{PRB-NENB}

\begin{figure}
\includegraphics[width=0.45\textwidth,clip=true]{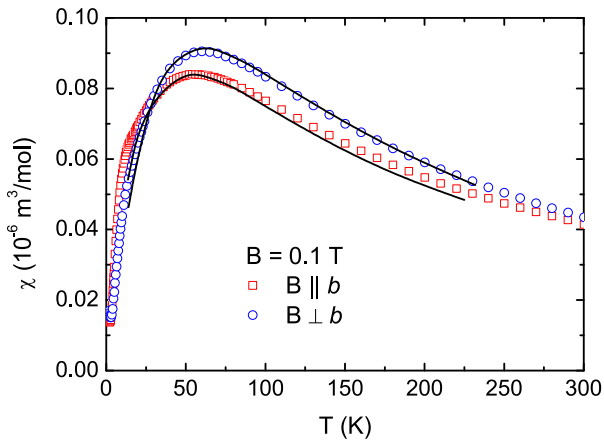}
\caption{\label{fig:susc1} (Color online) Magnetic susceptibility
of NENB measured in a magnetic field of 0.1~T, applied along
(squares) and perpendicular (circles) to the $b$ axis. The solid
lines represent calculations using the $S=1$ Heisenberg AFM
chain model with $D/J=0.2$ (see the text for details).}
\end{figure}

The static susceptibility was measured using a commercial
SQUID magnetometer equipped with a 7-Tesla magnet in the
temperature range from 1.8 to 300~K. The sample was attached with
a small amount of Apiezon N grease to the inside of a straw held
by a sample-holder rod. The core diamagnetic contribution to the
magnetic moment of the sample, calculated using Pascal's
constants,\cite{Kahn} was subtracted from the raw data. The
magnetization was measured in fields up to 15~T using a torque
magnetometer in a dilution refrigerator. ESR experiments were
performed in the Voigt geometry with the external field applied
along the $b$ axis using a tunable-frequency ESR setup at the
National High Magnetic Field Laboratory, Tallahassee.
\cite{spectrometer}

\begin{figure}
\includegraphics[width=0.45\textwidth,clip=true]{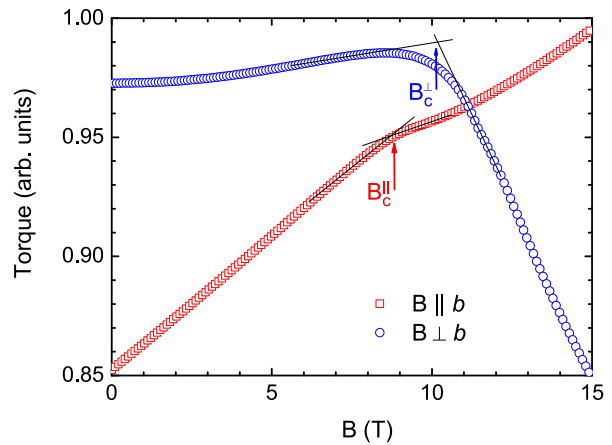}
\caption{\label{fig:torque} (Color online) Field dependence of the
magnetic torque of NENB at 20~mK for two field orientations.}
\end{figure}

The zero-field-cooled magnetic susceptibility of NENB measured
in a field of $B=0.1$~T applied parallel and perpendicular to
the $b$ axis is shown in Fig.\ \ref{fig:susc1}. The pronounced
maximum and low-temperature behavior of the magnetic
susceptibility suggest a spin-singlet ground state, which is a characteristic 
property of $S=1$ Haldane systems. Due to anisotropy effects, the
position of the susceptibility maximum depends on the field
orientation. The existence of a finite crystal-field anisotropy in
NENB has been confirmed by low-temperature (20~mK) magnetization
measurements using the torque magnetometry technique. Two critical fields
($B^{\parallel}_{c}\sim9$~T and $B^{\perp}_{c}\sim10$~T for $B$ applied parallel and perpendicular
to the $b$ axis, respectively) were resolved in the magnetization
data (Fig.\ \ref{fig:torque}), suggesting a field-induced collapse
of the spin-singlet ground state for both orientations of the magnetic field. The critical fields revealed in our experiments agree well with those obtained from optical and
magnetization measurements.\cite{PRB-NENB}

\begin{figure}
\includegraphics[width=0.45\textwidth,clip=true]{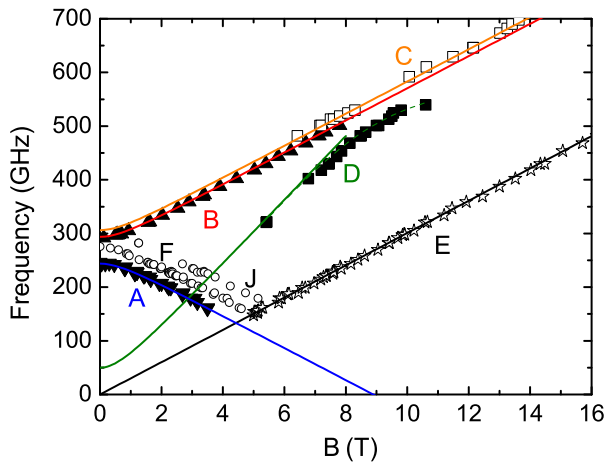}
\caption{\label{fig:epr} (Color online) Field dependence of the
magnetic ESR excitations of NENB observed at 1.5~K with the
magnetic field aligned along the $b$ axis. Solid lines represent
the fit of effective Hamiltonian (Eq.\ \ref{effham}) to the
experimental data.}
\end{figure}

To get a deeper insight into the peculiarity of magnetic interactions in NENB,
tunable-frequency ESR measurements have been performed in magnetic fields
up to 16 T. The excitation spectrum of NENB in fields parallel to
the $b$ axis (which is the chain axis) is plotted in Fig.\ \ref{fig:epr}. Several ESR modes
have been observed. In order to assign these modes to
field-dependent transitions in the energy scheme of magnetic excitations we analyzed the
data using the effective Hamiltonian, written as\cite{effHam}
\begin{eqnarray}
{\cal H}=\Delta+D'(S^z_i)^2+E'[(S^x_i)^2-(S^y_i)^2]-\mu_BSgB
\label{effham}
\end{eqnarray}
with eigenstates of type $\left|S,S^{z}\right\rangle$, where
$\Delta$ is the Haldane gap, $D'$ and $E'$ represent reduced
parameters of the crystal field anisotropy (uniaxial- and rhombic-distortion parameters, respectively) of the Ni ions, while the last term is
the Zeeman energy for an ion with $S=1$.

\begin{figure}
\includegraphics[width=0.45\textwidth,clip=true]{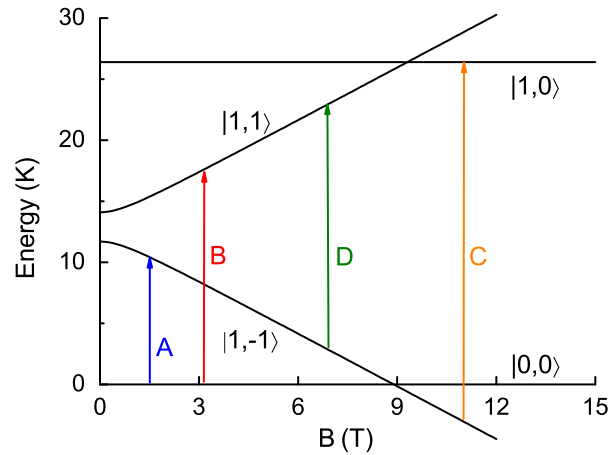}
\caption{\label{fig:energy} (Color online) Proposed
field-dependent energy scheme of NENB (using Eq.\ \ref{effham})
for magnetic fields applied along the $b$ axis.}
\end{figure}

The field-dependent energy scheme (Fig.~\ref{fig:energy})
corresponding to this Hamiltonian allows one to understand most of
the observed ESR modes. First, the transitions from the ground
state to the first excited states ($\left|0,0\right\rangle
\rightarrow\left|1,\pm1\right\rangle$), denoted by A and B, have
been observed in fields up to $\sim$~8~T as shown by the triangles in
Fig.\ \ref{fig:epr}. The extrapolation of the mode A to zero
frequency suggests the existence of a critical field $B_c^
{\parallel}\approx9$~T, which coincides nicely with the change of
the torque signal for fields applied parallel to the $b$ axis (Fig.\ \ref{fig:torque}). At
this field, the first excited level crosses the nonmagnetic ground
state transforming the system into a state with a finite
magnetization. The transitions A and B would be forbidden in a system with purely axial symmetry, but can be allowed in the presence of
a staggered magnetization.\cite{stagger1} As suggested by
Sakai and Shiba\cite{stagger2}, the staggered term lifts
wave-vector selection rules and allows transitions from the ground
states to the excited states at the boundaries of the Brillouine zone. The staggered magnetic moment may originate from an alternating $g$ tensor (which was revealed by crystallographic analysis of NENB) or a finite Dzyaloshinskii-Moriya interaction.

The mode D corresponds to transitions $\left|1,-1\right\rangle
\rightarrow\left|1,1\right\rangle$, which are only allowed in the
presence of the broken axial symmetry. As follows from our ESR data, the excited doublet is split in zero magnetic field, which is a clear signature of a finite rhombic distortion. The mode C corresponds to
$\left|1,-1\right\rangle\rightarrow\left|1,0\right\rangle$
transitions observed in fields above $\sim$~6~T. The observation of this excitation mode is of
particular importance, since it gives direct evidence for the four-level
excitation scheme, which is a typical feature of an $S=1$ Haldane-chain
system with finite anisotropy. Then, in accordance to Eq.\ \ref{effham} for the Haldane gap we found $\Delta=17.4$~K.

Two more resonance absorptions (the modes F and J, Fig.\ \ref{fig:epr}) have been
observed in the low-temperature excitation spectrum of NENB. The
field dependences of these excitations are similar to that of mode
A. The excitation energies are higher by 30 and 60 GHz, for the modes F and J,
respectively. The origin of these
excitations is not clear at the moment. As mentioned above, the
orientation of the nitrite ions within a single chain appears to
be uniform (one N atom and one O atom are coordinated to Ni$^{2+}$
center along the chain direction). On the other hand, as mentioned in Ref.~\onlinecite{structNENB}, one can not
exclude the possibility of two additional, minor Ni$^{2+}$ sites within the chains caused by the reversal of the orientation of the NO$_{2}^{-}$ groups. This would lead to occasional positions where a Ni$^{2+}$ ion in NENB would have either two O atoms from nitrite ions, or two N atoms from nitrite ions. Such a possibility was also suggested for the isostructural perchlorate compound NENP.\cite{structNENP} It is important to mention here that additional states close to the 
Haldane gap were observed in NENP by means of high-field ESR,\cite{LuNENP,SielingNENP} which is fully consistent with our observations.   
Another possible explanation of the extra states observed by us is the presence
of effective ferromagnetic impurity-induced bonds as revealed by susceptibility measurements, see discussion below. Such bonds were found to be responsible for the splitting of the neutron scattering
peak into two incommensurate parts placed symmetrically around the boundary of the Brillouine zone close to the gap energy in Y$_{2-x}$Ca$_x$BaNiO$_5$.\cite{impbond} However, it is fair to mention
that to prove this mechanism to be present in NENB requires more
detailed investigations.

Finally, a strong resonance line E, with $g=2.14$
was observed. The observation of the mode suggests the presence of spin-1/2
states (which origin can be ascribed to fractional chain-end effects or multiply frustrated interchain
interactions) even in nominally pure samples of NENB. These effects, as suggested below, appear to affect the low-temperature magnetic susceptibility behavior.

Using the Hamiltonian Eq.\ \ref{effham}, the ESR data can be fit with parameters 
$\Delta=17.4$~K, $D'=-13.5$~K, $\left|E'\right|=1.2$~K, and
$g_{\parallel}=2.14$. Numerical calculations for the
diagonalization of the Hamiltonian of $S=1$ anisotropic
exchange-coupled spin-chain system
\begin{eqnarray}
{\cal
H}=-J\sum_i{S_iS_{i+1}}+\sum_i{\left\{D(S^z_i)^2+E[(S^x_i)^2-(S^y_i)^2]\right\}}\nonumber
\\-\sum_i{\mu_BS_igB},~ \label{fullham}
\end{eqnarray}
outlined in Ref.~\onlinecite{LuNENP} suggest the relations
$D=-D'/1.8$ and $\left|E\right|=\left|E'\right|/1.7$, which yield
crystal-field anisotropy parameters $D=7.5$~K and
$\left|E\right|=0.7$~K.

Using the obtained set of parameters, the magnetic susceptibility has
been fit by the high-temperature series expansion for the $S=1$
AFM spin-chain model\cite{Weng,structNENP} yielding $J=-44.8~K$,
$g_\parallel =2.15$ and $J=-47.7~K$, $g_\perp =2.28$ for
$B\parallel b$ and $B\perp b$, respectively. The relatively strong
crystal-field anisotropy, $D=7.5$~K, as estimated from our ESR data,
yields a ratio $D/\left|J\right|=0.17$. This suggests that another model (where
the influence of anisotropy is taken into account) might be more
appropriate for describing the magnetic susceptibility.\cite{anis}
Although theoretical predictions are available for $D/\left|J\right|=0.2$, the
overall agreement with the experimentally determined
susceptibility data is very good (see Fig.\ \ref{fig:susc1}). We obtain
$J=-45$~K, $g_\parallel=2.15$ and $J=-46.5$~K, $g_{\perp}=2.28$
for $B\parallel b$ and $B\perp b$, respectively. Combining these
values of the exchange coupling with the ESR results for $\Delta$
we obtain $\Delta/\left|J\right|$ in the range from 0.39 to 0.42, which is in
good agreement with theoretical predictions.\cite{DMRGgap,
exactgap}

\begin{figure}[t]
\includegraphics[width=0.45\textwidth,clip=true]{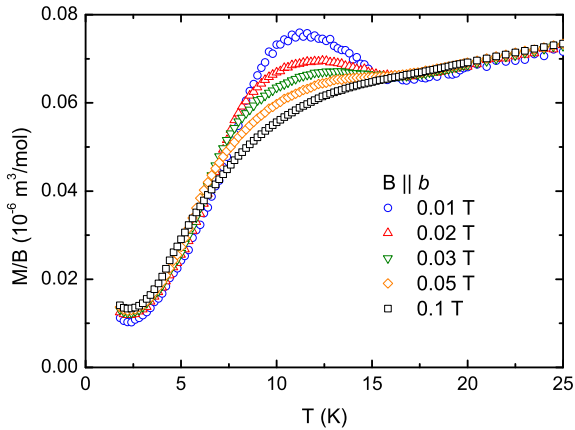}
\caption{\label{fig:susc2} (Color online) Temperature dependence
of the magnetic susceptibility of NENB in different magnetic
fields applying along the chain $b$ axis.}
\end{figure}

The presence of fractional $S=1/2$ chain-end spins might be a
possible explanation for the observation of the resonance line E
in the ESR spectrum as well as for a weak Curie-like
low-temperature upturn observed in the susceptibility. The
low-temperature susceptibility measured in magnetic fields from
0.01 to 0.1~T (applied parallel to the $b$ axis) is shown in Fig.\
\ref{fig:susc2}. The susceptibility is characterized by a small
but clearly resolvable local maximum at $\sim12$~K, which gets
rapidly suppressed with increasing magnetic field. Such behavior
can be accounted for by the formation of a spin-glass state that
has been also observed in powdered NENB samples. \cite{spinglassNENB}

When structural defects or non-magnetic impurities are introduced
in the spin chain, the VBS model predicts the presence of two
unpaired $S=1/2$ spins, one at each end of the open chain. The
existence of these end-chain states has been confirmed
numerically (using the Monte Carlo\cite{MCendchain} and DMRG%
\cite{Whiteendchain} calculations) as well as experimentally in
the Haldane material NENP doped with non-magnetic
ions.\cite{GlarumNENP} A pronounced signature of the $S=1/2$
chain-end interaction is a spin-glass behavior, as observed by
Hagiwara \textit{et al.}, \cite{spinglass1,spinglass2} who
proposed a model where structural defects induce ferromagnetic
bonds between the moments at the ends of finite chains.
On the other hand, the structural defects would allow a stronger interchain
coupling at the defect sites, which assuming a random
distribution of defects appears to induce frustrations and the spin-glass
behavior. Additional impurity-induced energy levels were calculated and observed in the quasi-one-dimensional Heisenberg-chain compound Y$_{2-x}$Ca$_x$BaNiO$_5$.\cite{midgap,impbond}

In conclusion, a systematic study of the new spin-1 AFM Heisenberg-%
chain system NENB has been presented. We showed conclusively, that
NENB belongs to the Haldane class of materials, with the energy
gap $\Delta=17.4$~K. Based on the analysis
of the ESR and magnetization data, the spin-Hamiltonian parameters
have been estimated, yielding the exchange coupling $J=-45$~K and
crystal-field anisotropy parameters $D=7.5$~K and $|E|=0.7$~K. Extra states in the vicinity
of the Haldane gap were observed, which origin is discussed. The presence of
fractional $S=1/2$ chain-end moments, revealed by means of ESR
and magnetization measurements, is found to be responsible for
spin-glass freezing effects, which is consistent with the VBS
model for $S=1$  Heisenberg AFM chains.

The authors would like to thank I. Affleck for fruitful
discussions. Part of this work was supported by the DFG (project No. ZV 6/1-1). Part of this work was performed at the
National High Magnetic Field Laboratory, Tallahassee, USA, which
is supported by NSF Cooperative Agreement No. DMR-0084173, by the
State of Florida, and by the DOE. S.A.Z. acknowledges the support from the NHMFL through the
VSP No. 1382. 

\bibliography{nenb}

\end{document}